\documentclass[%
 reprint,    
superscriptaddress,
 amsmath,amssymb,
 aps,
prb,
]{revtex4-2}

\usepackage{graphicx}
\usepackage{dcolumn}
\usepackage{bm}
\usepackage{xcolor} 

\begin{document}


\title{Non-volatile Orbital magnetoelectric effect in a two-dimensional electron gas}

\author{Yu Chen}
\email{yu.chen@spin.cnr.it}
\affiliation{CNR-SPIN, Complesso di Monte S. Angelo,
	Via Cinthia, 80126 Napoli, Italy}

\author{Martando Rath}
\affiliation{CNR-SPIN, Complesso di Monte S. Angelo,
	Via Cinthia, 80126 Napoli, Italy}

\author{Julien Brehin}
\altaffiliation{Present address: Univ. Grenoble Alpes, CEA, CNRS,
	Grenoble-INP, SPINTEC, 38000 Grenoble, France}
\affiliation{Laboratoire Albert Fert, CNRS, Thales,
	Université Paris-Saclay, 91767 Palaiseau, France}

\author{Cinthia Piamonteze}
\affiliation{Photon Science Division, Paul Scherrer Institut,
	5232 Villigen PSI, Switzerland}
	
\author{Daniele Preziosi}
\affiliation{Université de Strasbourg, CNRS, IPCMS UMR 7504, 67034 Strasbourg, France}

\author{Payal Wadhwa}
\affiliation{Department of Applied Physics,
	Indian Institute of Information Technology Vadodara,
	Gandhinagar 382028 Gujarat, India}
	
\author{Alessio Filippetti} 
\affiliation{Dipartimento di Fisica, Università di Cagliari, Cittadella Universitaria, 09042 Monserrato (CA), Italy}
\affiliation{CNR-IOM, Cagliari, Cittadella Universitaria, 09042 Monserrato (CA), Italy}

\author{Yuri Fukaya}
\affiliation{CNR-SPIN, Fisciano (Salerno), Italy and Dipartimento di Fisica “E. R. Caianiello,” Universitá di Salerno, 84084 Salerno, Italy}
\affiliation{Faculty of Environmental Life, Natural Science and Technology, Okayama
University, 700-8530 Okayama, Japan}

\author{Mario Cuoco}
\affiliation{CNR-SPIN, Fisciano (Salerno), Italy and Dipartimento di Fisica “E. R. Caianiello,” Universitá di Salerno, 84084 Salerno, Italy}

\author{Filomena Forte}
\affiliation{CNR-SPIN, Fisciano (Salerno), Italy and Dipartimento di Fisica “E. R. Caianiello,” Universitá di Salerno, 84084 Salerno, Italy}

\author{Daniela Stornaiuolo}

\affiliation{Dipartimento di Fisica, Università
	``Federico II'' di Napoli, Via Cinthia, 80126 Napoli, Italy}
	
\affiliation{CNR-SPIN, Complesso di Monte S. Angelo,
	Via Cinthia, 80126 Napoli, Italy}

\author{Manuel Bibes}
\affiliation{Laboratoire Albert Fert, CNRS, Thales,
	Université Paris-Saclay, 91767 Palaiseau, France}

\author{Marco Salluzzo}
\email{marco.salluzzo@spin.cnr.it}
\affiliation{CNR-SPIN, Complesso di Monte S. Angelo,
	Via Cinthia, 80126 Napoli, Italy}

\date{\today}

\begin{abstract}
Electrical control of orbital magnetism is a central goal of orbitronics. While current-induced orbital polarization has recently been demonstrated in inversion-asymmetric systems, nonvolatile electrical control of an equilibrium orbital moment has remained elusive.

Here, using x-ray absorption spectroscopy with \textit{in situ} electric-field control of the FE polarization, we demonstrate non-volatile switching of the equilibrium orbital magnetization of the oxide q2DEGs formed at the LaAlO$_3$/EuTiO$_3$/Ca-STO (001) interface. X-ray linear dichroism reveals a reversible modulation, driven by the FE-polarization direction, of the Ti-3$d$ crystal-field splitting and orbital occupancy, establishing a ferroelectric-like metallic state. 
X-ray magnetic circular dichroism shows a polarization-controlled switching of the  Eu$^{2+}$ orbital magnetic moment induced by the orbital magnetization of the q2DEG projected onto Eu-5$d$ states and detected at the Eu-$M_{4,5}$ edge through Eu-5$d$/Eu-4$f$ hybridization in the inversion-symmetry-broken interface.
These results provide spectroscopic evidence of ferroelectric control of orbital magnetism in an oxide q2DEG.
\end{abstract}

\maketitle


\section{\label{sec:level1} Introduction}

Since their discovery \cite{Ohtomo2004}, oxide quasi-two-dimensional electron gases (q2DEGs) have revealed a remarkably rich phenomenology, fostering intense interest for all-oxide electronic and spintronic platforms. The LaAlO$_3$ (LAO)/SrTiO$_3$ (STO) q2DEG exhibits a variety of fascinating properties, including an electric-field-induced insulator-to-metal transition \cite{Thiel2006}, gate-tunable Rashba spin-orbit coupling \cite{Caviglia2010,Stornaiuolo2014}, and unconventional superconductivity \cite{Caviglia2008,Richter2013,Stornaiuolo2017}. More recently, oxide q2DEGs have emerged as promising platforms for orbitronics owing to their sizable orbital textures and electrically tunable orbital angular momentum (OAM) \cite{Johansson2021,ElHamdi2023}, while simultaneously exhibiting efficient spin-to-charge conversion \cite{Lesne2016}.

In this context, realizing oxide q2DEGs that combine ferroelectric (FE) and ferromagnetic (FM) order parameters \cite{Brehin2023} is expected to greatly expand their functional landscape by enabling non-volatile electrical control of both spin and orbital degrees of freedom, potentially addressing one of the outstanding challenges in orbitronics: the realization of q2DEG devices whose equilibrium orbital magnetization can be switched purely by an electric field.

Coexisting and coupled functionalities can be engineered at oxide interfaces through epitaxial design. In particular, spin-polarized q2DEGs can be realized by inserting a thin magnetic layer between LAO and STO, such as EuTiO$_3$ (ETO) \cite{DeLuca2014,Stornaiuolo2016} or LaMnO$_3$ \cite{Zhang2017}, or by replacing LAO with a ferro(ferri)magnetic insulator, as demonstrated in GdTiO$_3$/STO \cite{Moetakef2011}, EuO/STO \cite{Kormondy2018,rosenberger2024,lazrak2025}, and LaAl$_{1-x}$Mn$_x$O$_3$/STO \cite{Yulin2019}. 

Conversely, the realization of genuine FE metals appears paradoxical, since itinerant carriers efficiently screen electric dipoles, suppressing long-range FE order. Nevertheless, Anderson and Blount proposed that a FE-like transition can, in principle, occur in metals \cite{Anderson1965}. This concept has gained renewed momentum with the discovery of two-dimensional FE-metals, such as WTe$_2$ \cite{Fei2018}, layered oxides Bi$_n$Ti$_n$O$_{2n+5}$ \cite{Filippetti2016} and Bi$_n$Mn$_n$O$_{2n+5}$ \cite{Urru2020}, and oxide 2DEGs formed at interfaces between LaAlO$_3$ or AlO$_x$ thin films and bulk FE Ca-doped STO (Ca-STO) single crystals \cite{Brehin2020}. In the latter system, the FE transition and the bulk Ca-STO polarization strongly modulate interfacial transport properties \cite{Brehin2020}, opening a promising route toward non-volatile, low-power electronic and spin-orbitronic devices controlled by electric fields.

Recently, we realized an oxide q2DEG combining both FE and FM order parameters by sequential deposition of ETO and LAO on TiO$_2$-terminated Ca(1\%)-doped STO \cite{Brehin2023}. Magnetotransport measurements revealed a coupling between ferroelectricity and magnetism, while Ti-$L_{2,3}$ and Eu-$M_{4,5}$ x-ray magnetic circular dichroism (XMCD) confirmed the formation of a FM q2DEG, consistent with anomalous Hall effect measurements. 

Here, we provide spectroscopic evidence for non-volatile electric-field switching of the equilibrium orbital magnetization associated with the LAO/ETO/Ca-STO q2DEG, enabled by the FE-control of its electronic and magnetic properties. Using x-ray absorption spectroscopy (XAS) with \textit{in situ} back-gate voltage control, we demonstrate that the system behaves as a ferrolectric two-dimensional metal, exhibiting a pronounced modulation of Ti-3$d$ orbital splitting and occupancy across the FE transition and as a function of the polarization direction. More importantly, we show a clear non-volatile modulation of the Eu$^{2+}$ orbital (and spin) moment with gate voltage. By combining a multi-orbital tight-binding model, derived from density functional theory, with a configuration-interaction model of the Eu $4f$ and $5d$ states, we show that the orbital magnetization carried by the q2DEG is transferred from Ti-3$d$ states to Eu-5$d$ orbitals and becomes detectable at the Eu-$M_{4,5}$ edge through enhanced Eu-5$d$/Eu-4$f$ hybridization. These results highlight the potential of artificial oxide q2DEGs as a platform for orbitronic devices with non-volatile electrical control of the equilibrium orbital magnetization.

\begin{figure}[ht!]
	\includegraphics{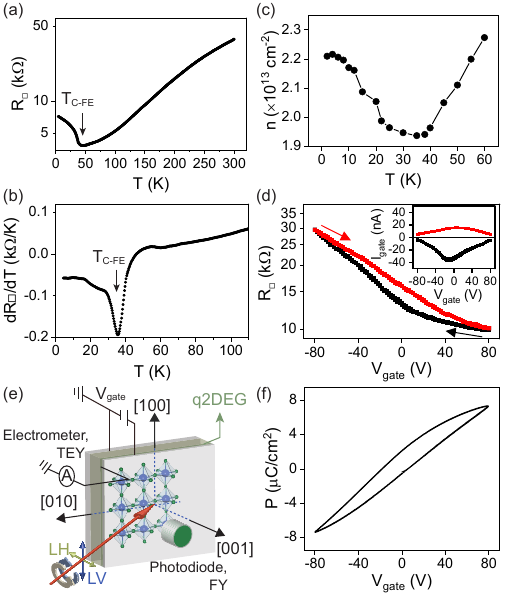}
	\caption{\label{fig1} (a) Temperature dependence of the sheet resistance and (b) of its derivative respect the temperature of a LAO/ETO/Ca(2\%)-STO sample, showing the FE-Tc (black arrows). (c) Temperature dependence of the sheet carrier density across the FE-transition.  (d) Sheet resistance vs. back-gate voltage at 4K showing non-volatile modulation of the transport properties. The gate-current vs. the gate voltage (inset), measured at the same time, shows a typical FE-switching behaviour.  (e) Schematic of the experimental set-up and geometry for the XLD and XMCD experiment under application of a back gate voltage. (e) Polarization vs. V\textsubscript{gate} on LAO/ETO/Ca-STO (black) measured during the XAS experiment.}
\end{figure}

\section{\label{sec:level2} Results}
Heterostructures composed of 6 unit cells of LAO and 2uc of ETO were grown by pulsed laser deposition (PLD) on Ca(2\%) doped-STO(001) single crystals (see Supplemental Material \cite{SI}). The resulting q2DEGs exhibited transport properties comparable to state-of-the-art Ca–STO–based interfaces. As shown in Figs.~\ref{fig1}(a,b), the onset of ferroelectricity is marked by an increase in the sheet resistance below $\sim$40 K, with the maximum slope occurring near the FE transition temperature $T_{c-FE}$ $\approx$ 35 K. The carrier density at 2 K is $\sim$2.2 $\times$ 10$^{13}$ cm$^{-2}$, comparable to conventional LAO/STO and LAO/ETO/STO interfaces, exhibits a pronounced anomaly across $T_{c-FE}$ (Fig. \ref{fig1}c). 

Polarization switching at low temperature is achieved by applying gate voltages between 60–80 V applied to a metallic back-gate deposited on the 0.5-mm-thick Ca-STO substrate. The gate-voltage dependence of the sheet resistance at 4 K (Fig. \ref{fig1}d) displays a pronounced hysteresis associated with the FE-switching,  while the simultaneously measured gate current (inset of Fig.~\ref{fig1}d) exhibits the corresponding polarization-switching peaks. The P$\uparrow$ state increases both the carrier density and the sheet conductance, whereas the P$\downarrow$ state produces the opposite effect. Importantly, the $R_{sheet}$ hysteresis direction is opposite to that expected from charge trapping and de-trapping effects commonly observed in oxide q2DEGs \cite{Lyzwa2022}, demonstrating that the transport response is governed by the FE-polarization.

The experimental configuration used to probe the effect of the Ca–STO polarization during the XAS experiment is schematically illustrated in Fig. \ref{fig1}e. The back-gate electrode was connected through an electrically isolated coaxial cable, while the sample was mounted on a 100 $\mu$m thick sapphire plate providing both electrical insulation and excellent thermal anchoring at low temperature. The q2DEG was electrically contacted to the sample holder using aluminum bonding wires and silver paint, allowing TEY detection, as well as \textit{in situ} FE-characterization, and application of a dc back gate (V$_{gate}$). A representative polarization vs. V$_{gate}$ loop acquired during the XAS experiment is shown in Fig. \ref{fig1}f.

Unless otherwise stated, all the spectroscopy results presented in this work were acquired under zero applied gate voltage, after setting the sample in a well-defined non-volatile upward (P$\uparrow$) or downward (P$\downarrow$) polarization state. After each spectroscopic measurement, the ferroelectric state was verified by recording a new polarization loop before switching to the opposite polarization. This procedure was repeated several times throughout the experiment, confirming the stability and reproducibility of the non-volatile polarization states.

Polarization-dependent XAS at the Eu–M$_{4,5}$ and Ti–L$_{2,3}$ edges was performed at the X-Treme beamline of the Swiss Light Source \cite{Piamonteze2012}. XAS spectra were collected in both total electron yield (TEY) and total fluorescence yield (TFY) modes. 
The interfacial Ti-$3d$ orbital splitting and magnetism was investigated by Ti–L$_{2,3}$ edge XLD and XMCD measurements carried out in TEY mode (probe depth $\approx$ 2 nm) under grazing-incidence ($\theta=60^\circ$).

The magnetic properties of Eu-$4f$ were investigated by Eu–M$_{4,5}$ XMCD. Eu-$M_{4,5}$ XMCD measurements were carried out in TFY mode. Since Eu ions are confined to the thin 2uc interfacial EuTiO$_3$ layer, self-absorption effects are negligible. Moreover, TFY detection is insensitive to gate-induced changes in the sample drain current that might affect TEY measurements, enabling reliable XMCD acquisition as a function of the back-gate voltage along the P–V\textsubscript{gate} hysteresis loop. The Eu-$M_{4,5}$ edge probes predominantly the Eu-$4f$ magnetic moment through the dipole-allowed $3d\rightarrow4f$ transition.

\subsection{Experimental evidence of FE modulation of Ti–3d states}
To investigate the electronic fingerprints of ferroelectricity at the interface, we performed Ti-L$_{2,3}$ XLD measurements across the FE-transition as function of the ferroelectric polarization states.
Ti-L$_{2,3}$ XLD spectra, measured in the TEY mode, directly probes the orbital splitting and anisotropic occupation of Ti 3d t$_{2g}$ and e$_{g}$ orbitals, as previously demonstrated for LAO/STO heterostructures \cite{Salluzzo2009}. 

\begin{figure}
	\includegraphics[width=8.0 cm]{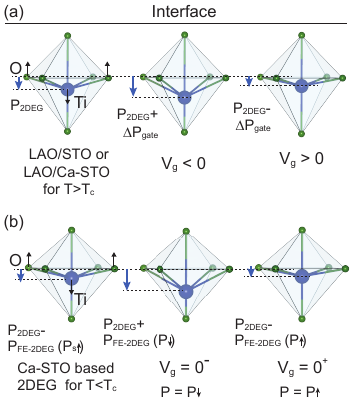}
	\caption{\label{fig2} (a) Sketch of the effect of the gate on the Ti cations and oxygen ions  displacements as a function of an external back-gate voltage in a standard LAO/STO system and paraelectric Ca-STO based q2DEG above Tc.(b) Sketch of the effect of the polarization in the unpoled state (spontaneous polarization), P$\downarrow$ ($V_g=0^-$) and P$\uparrow$ ($V_g=0^+$) states.}
\end{figure}

In STO- and Ca-STO- based q2DEGs above the $T_{c-FE}$, the electrostatic confining potential generated by the interfacial charge transfer induces polar distortions of the Ti$O_{6}$ octahedra, with Ti-ions displaced towards the substrate and oxygen ions towards the interface (Fig. \ref{fig2}a). These Ti-O displacements generate an interfacial electrical dipole that lifts the orbital degeneracy of Ti-$3d$ states and gives rise to the characteristic XLD signal. In LAO/STO, the Ti-$3d$ crystal-field splitting changes sign when the LAO thickness exceeds the critical value of four unit cells, coinciding with the onset of electronic reconstruction and q2DEG formation \cite{Salluzzo2009,Savoia2009}.
The resulting interfacial dipole can be continuously tuned by an external gate voltage through changes in the confining potential. However, because STO is not ferroelectric, this modulation is volatile and disappears once the electric field is removed.

\begin{figure}
	\includegraphics[width=8.0 cm]{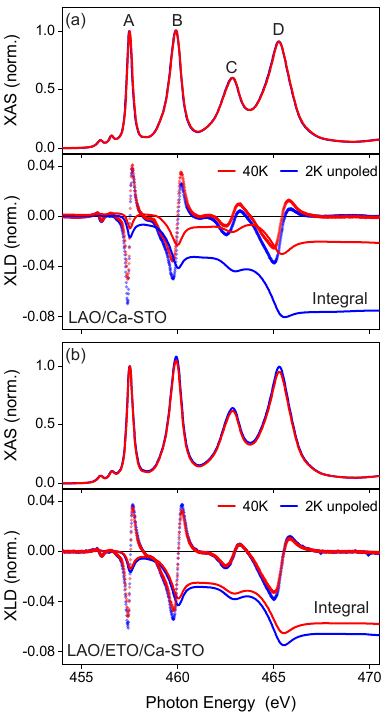}
	\caption{\label{fig3} (a)-(b) Ti L$_{2,3}$ XAS and XLD measurements across the FE-transition on (a) LAO/Ca-STO and (b) LAO/ETO/Ca-STO. Scatter red and blue data in the XLD panels are obtained just above (40 K) and well-below (2K) the FE-Tc, respectively, while the corresponding continuous lines are the XLD integrals.}
\end{figure}

\begin{figure}
	\includegraphics[width=8.0 cm]{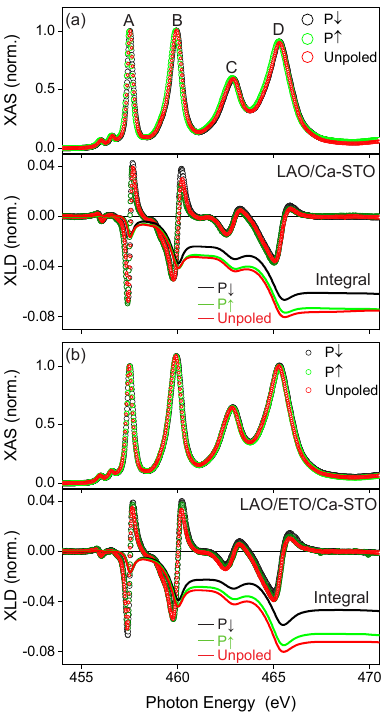}
	\caption{\label{fig4} (a)-(b) Ti L$_{2,3}$ XAS and XLD measurements as function of the FE-polarization direction on (a) LAO/Ca-STO and (b) LAO/ETO/Ca-STO. In the XLD panels, data (scatter) and integrated XLD (continuous lines) at 2K  were measured in the P$\downarrow$ (black), P$\uparrow$ (green) and spontaneous polarization (unpoled) state (red).}
\end{figure}

In contrast, when STO is replaced by ferroelectric Ca-STO, the spontaneous polarization of the substrate couples to the interfacial dipole of the q2DEG. As a consequence, the Ti-$3d$ orbital splitting and orbital occupation become non-volatile quantities that retain the memory of the ferroelectric polarization state even after removal of the external electric field, as already demonstrated in ref.~\cite{Brehin2023}. This corresponds to the presence of a ferroelectric polarization component, P$_{FE-2DEG}$, due to the 2DEG itself (Fig. \ref{fig2}b), that can be switched upward or downward with the back-gate. 

Figures \ref{fig3}(a,b) show the Ti-$L_{2,3}$ XAS and XLD spectra of LAO/Ca-STO and LAO/ETO/Ca-STO interface measured above and below the ferroelectric transition temperature of Ca-STO. As in conventional LAO/STO q2DEGs, the overall XLD line shape is dominated by the dichroic response of Ti$^{4+}$ ($3d^0$) ions and reflects the crystal-field splitting of the Ti-$3d$ states. The spectra indicate that both the $t_{2g}$ and $e_g$ manifolds are split with the in-plane orbitals, $3d_{xy}$ and $3d_{x^2-y^2}$, lying at lower energy than out-of-plane $3d_{xz,yz}$ and $3d_{z^2}$ orbitals. Accordingly, $\Delta t_{2g}$ and $\Delta e_{g}$ crystal field splittings are negative.

In contrast, the integrated XLD signal probes the relative occupation of in-plane (n(3d$_{xy}$)) and out-of-plane (n(3d$_{xz,yz}$)) Ti$^{3+}$ ($3d^1$) states. According to the XLD sum rules
\begin{equation}
\dfrac{n(3d_{xy})}{n(3d_{xz,yz})} \cong  \dfrac{\int (I_c - I_{ab})dE}{\int (I_c + I_{ab})dE},
\end{equation}
so that a negative XLD integral corresponds to a preferential occupation of Ti$^{3+}$-3d$^{1}_{xz,yz}$ orbitals. It is worth noting that the XLD sum rule accounts for the total occupied Ti-$3d$ spectral weight, and includes mobile as well as localized electronic states. Thus, while the shape of the XLD indicates that $3d_{xy}$ states are lower in energy than $3d_{xz,yz}$ states, the negative XLD integral suggests the presence of localized electrons with Ti$^{3+}$-3d$^{1}_{xz,yz}$ character. 

The FE transition produces two distinct signatures in the XLD spectra. First, the dichroic amplitude increases over both t$_{2g}$ (A, C peaks) and e$_g$ (B, D peaks) regions in the two samples, indicating an enhancement of the crystal-field splitting. Second, the XLD integral becomes more negative below $T_{c-FE}$, demonstrating increased occupation of Ti$^{3+}$-$3d_{xz,yz}$ states.

\begin{table*}
\caption{\label{table1} Crystal field parameters $D_s$, $D_t$, charge transfer energy $\Delta = 2.5$ eV and hopping integrals T($a_1$), T($b_1$), T($a_2$), T(e) in the different polarization states and at different temperatures after fitting the XLD data (Fig.\ref{fig3} and Fig.\ref{fig4}) on LAO/Ca(2\%)-STO and LAO/ETO/Ca(2\%)-STO. The $t_{2g}$ and $e_g$ splitting, $\Delta t_{2g}$ and $\Delta e_g$, are determined from $D_s$ and $D_t$ parameters. The Slater integrals were reduced by 90\% with respect to the corrected Hartree-Fock values. A Gaussian broadening of 0.1 eV, related to the experimental energy resolution, and Lorentzian broadenings for each peak were applied to reproduce the experimental line-widths.}

\begin{ruledtabular}
\begin{tabular}{cccccccccc}
Sample & 10 $D_q$ & $\Delta t_{2g}$ & $\Delta e_g$ & $D_s$ & $D_t$ & T($a_1$) & T($b_1$) & T($a_2$) & T(e)\\
LAO/2\%Ca-STO & (eV) & (meV) & (meV) & (meV) & (meV) & (eV) & (eV) & (eV) & (eV) 
\\ \hline
P$\downarrow$ 2K & 2 & -35  & -140 & -25 & -8  & 2.15  & 2  & 0.8  & 1.04 
\\ \hline
P$\uparrow$  2K & 2 &  -55 & -120 & -25 & -4  & 2.1  &  2 & 0.87  &  1.04 
\\ \hline
P\textsubscript{unpol} 2K& 2 & -60  & -115  & -25 & -3  & 2.1  & 2  & 0.87 &  1.04 
\\ \hline
P\textsubscript{unpol} 40K & 2 &  -56 & -104 & -23  & -2.5& 2.05   & 2  & 0.87  &  1.04 
\\ \hline\hline
Sample & 10 $D_q$ & $\Delta t_{2g}$ & $\Delta e_g$ & $D_s$ & $D_t$  & T($a_1$) & T($b_1$) & T($a_2$) & T(e)\\
LAO/ETO/2\%Ca-STO & (eV) & (meV) & (meV) & (meV) & (meV) & (eV)  & (eV) & (eV) & (eV) 
\\ \hline
P$\downarrow$ 2K &2	&-41&-148&-27&-8&2.15&2	&0.8&1.035
\\ \hline
P$\uparrow$  2K &2& -48&-134&	-26	&-6&2.1&2&0.87&1.04
\\ \hline
P\textsubscript{unpol} 2K& 2 & -45	&-130&-25&	-6&	2.1	& 2&	0.87&	1.04
\\ \hline
P\textsubscript{unpol} 40K & 2 &  -59&	-102&	-23&	-2&	2&	2&	0.93&	1 \\ \hline
Error-bar & - & $\pm6$ & $\pm7$ & $\pm0.5$ & $\pm0.1$ &  $\pm0.05$ &  $\pm0.025$ &  $\pm0.025$ & $\pm0.025$ \\
\end{tabular}
\end{ruledtabular}
\end{table*}

The effect of \textit{in-situ} ferroelectric polarization switching at 2K is shown in Fig. \ref{fig4}(a). The XLD spectrum measured in the P$\uparrow$ state is nearly identical to the remnant (unpoled) state, suggesting that the interface is intrinsically characterized by a spontaneous P$\uparrow$ configuration in both LAO/Ca-STO and LAO/ETO/Ca-STO. Reversing the polarization to P$\downarrow$ produces clear spectral changes, particularly at e$_g$ related features B and D and, to a lesser extent, at $t_{2g}$ features A and C. Simultaneously, the XLD integral becomes less negative, revealing a reduced occupation of Ti$^{3+}$-3d$_{xz,yz}$ states. These results demonstrate that the ferroelectric polarization controls the orbital splitting and occupation of Ti-3d states in a non-volatile way. 

To quantify the effect of the FE-transition and of the polarization direction on the Ti-3$d$ crystal-field splitting, the XLD spectra were analyzed using ligand-field atomic multiplet calculations implemented in the CTM4XAS \cite{Stavitski2010} and Quanty \cite{Haverkort2012} codes. The calculations employed a TiO$_{6}$ cluster in D${4h}$ symmetry and included Ti-3$d$–O-2$p$ hybridization through a charge-transfer model with a 3$d^0$+3$d^1\underline{L}$ ground-state configuration (\underline{L} denotes a ligand hole). The charge-transfer energy was fixed to $\Delta=2.5$ eV and the Coulomb interaction difference to $U_{pd}-U_{dd}=2$ eV, while the hopping integrals were chosen close to typical values for nearly octahedral TiO$_{6}$ environments, namely $T(b_1)$, $T(a_1)\approx2$ eV and $T(b_2)$, $T(e)\approx1$ eV. A quantitative reproduction of the XAS and XLD spectra for the different temperature and polarization states required, on the other hand, small adjustments of the crystal-field and hopping parameters, indicating a departure from ideal octahedral symmetry in the FE-state (see Supplemental Material \cite{SI}).
The optimized fitting parameters, summarized in Table 1, reveal a clear temperature- and polarization-dependent modulation of both the $t_{2g}$ and $e_g$ crystal-field splitting. In particular, we do observe larger splitting between in-plane and out-of-plane $e_{g}$ orbitals below the FE-Tc, and a larger $\Delta e_{g}$ in P$\downarrow$ configuration compared to P$\uparrow$ in both systems. On the other hand, the onset of ferroelectricity does not reverse the sign of the interfacial confining potential, which would suppress the q2DEG, and the sign and the shape of the dichroism is similar to the one of the LAO/STO q2DEG. Instead, below the FE-T$_{c}$ Ti–O polar distortions within the conducting region increase (larger eg splitting), and they are modulated by the polarization direction giving rise to a non-volatile switching of the interfacial electric dipole, in qualitative agreement with previous study on LAO/ETO/Ca(1\%)–STO heterostructures \cite{Brehin2023}. In particular, the FE polarization component associated with the q2DEG (P$_{\mathrm{FE-2DEG}}$) reverses together with the bulk Ca–STO polarization, providing direct spectroscopic evidence that the conducting interface behaves as a ferroelectric-like metal.

Beyond modifying the $t_{2g}$ and $e_g$ crystal-field splitting, the ferroelectric transition also induces a redistribution of the Ti-3$d$ sub-band occupation. In the P$\uparrow$ state, the enhanced population of the 3$d_{xz,yz}$ orbitals, together with the transport data, indicates that the q2DEG enters the multiband regime across the Lifshitz transition, with both the 3$d_{xy}$ and 3$d_{xz,yz}$ bands contributing to the electronic states at the Fermi level. As discussed below, this multiband regime plays a key role in the emergence of electrically switchable orbital magnetization.

\begin{figure*}[ht!]
	\includegraphics[width=155mm]{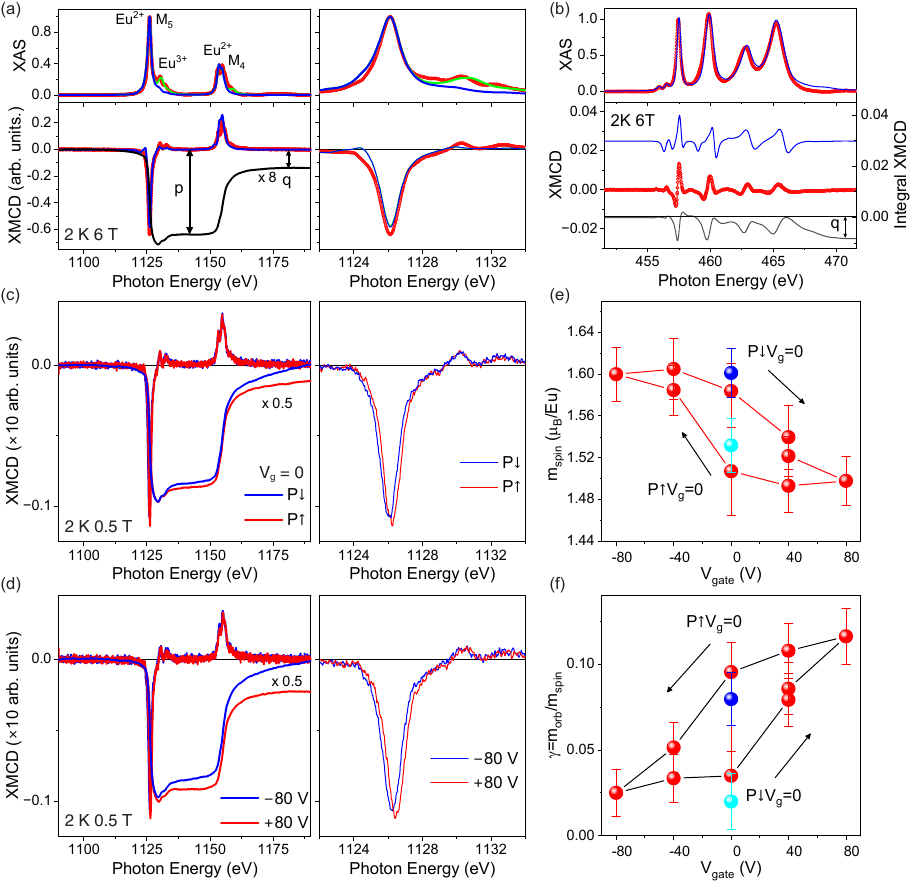}
	\caption{\label{fig5} (a) TFY Eu-M$_{4,5}$ edge XAS and XMCD data in saturation on the LAO/ETO/Ca-STO sample at 2 K and 6 Tesla (unpoled, spontaneous polarized state). In the upper panel we show the sum of the XAS in the circular polarizations (normalized to the maximum), while in the bottom panel the XMCD data normalized to M$_{5}$ XAS peak (red full circles). In the left panel we show the data on the full energy range, on the right we show the data around the M$_{5}$ peak. Blue and green lines are the calculated XAS and XMCD assuming 100\% Eu$^{2+}$ and 90\% Eu$^{2+}$+10\%Eu$^{3+}$ occupations, respectively. The XMCD calculated spectra with or without Eu$^{3+}$ are indistinguishable. Black continuous line in the bottom left panel is the integrated XMCD  (multiplied by a factor 8, see suppl. materials for details about background subtraction and data analysis).  p and q, are respectively the integral on the M$_5$ edge and the total XMCD integral. (b) TEY Ti-L$_{2,3}$ edge XAS (upper panel) and XMCD (bottom panel) data (red circles) and simulations (blue line) measured in the same conditions. In the bottom panel calculated and measured spectra, as well as the integrated XMCD (black-line), are displaced for clarity. (c) and (d) XMCD (thin lines) and integrated  (thick lines, scaled by a factor 0.5) XMCD data in the P$\uparrow$ (red) and P$\downarrow$ (blu) states at (c) Vg=0 and (d) Vg=$\pm$80 V (+80 V P$\uparrow$, -80V P$\downarrow$). (e) effective  spin-moment and (f) orbital over spin moment ratio as function of the back gate voltage, across the FE-loop. Blue scatter data (P$\uparrow$) and cyan scatter data (P$\downarrow$) in (e) and (f) were acquired with higher statistics.}
\end{figure*}

\subsection{Magnetism and Experimental evidence of FE-modulation of the Eu$^{2+}$ magnetic moment.}

The magnetic properties of LAO/ETO/Ca(2\%)-STO heterostructures have been investigated by Ti-L$_{2,3}$- and Eu M$_{4,5}$- edge XMCD. While interfacial Ti-magnetism can be single-out from the bulk signal only using TEY, TFY-Eu M$_{4,5}$-edge XMCD is better suited to probe the Eu-magnetism since it is less sensitive to interfacial defects.
Fig. \ref{fig6}a,b  show TFY- Eu M$_{4,5}$ and TEY- Ti L$_{2,3}$ edges XAS/XMCD data, acquired at 2K and 6 Tesla in grazing incidence (60 degrees from surface normal) on the LAO/ETO/Ca(2\%)-STO q2DEG in the spontaneous polarized state (unpoled). TFY-Eu M$_{4,5}$ spectra are dominated by the contribution from Eu$^{2+}$ ions, with a minor 10\% Eu$^{3+}$ fraction showing up only in the XAS spectra (Fig \ref{fig5}a), due to Eu$^{3+}$ replacing La$^{3+}$ in LAO at the LAO/ETO interface. On the other hand, the Eu$^{3+}$ contribution to the XMCD is negligible, as confirmed from identical XMCD spectra calculated assuming full Eu$^{2+}$ or a weighted average of Eu$^{2+}$ and Eu$^{3+}$ terms. Application of the 4f-sum rules yields a saturated Eu$^{2+}$ spin moment ($>$ 2 T) of approximately 6.4 $\mu_B$/Eu, not far from the value expected for Eu$^{2+}$ (4f$^7$, S = 7/2). The slightly reduced value is attributed to the 10\% Eu$^{3+}$ in the LAO layer due to intermixing.

It is worth noting, on the other hand, that the XMCD data at M$_{5}$- and at the M$_{4}$- edges sensitively depart from the simulated XMCD. As a consequence, the integrated experimentally measured XMCD is different from zero, revealing a very unusual sizeable orbital contribution corresponding to a m$_{orb}$/m$_{spin}$ ratio up to 0.1 and a calculated m$_{orb}$=0.6 $\mu_B$/Eu. While the quantitative application of the orbital-sum rules should be taken with some caution for Eu$^{2+}$, the non-vanishing orbital term and the large m$_{orb}$/m$_{spin}$ ratio is very unusual, as $<L>=0$ for Eu$^{2+}$ 4f states (for more details about the analysis of the XMCD data and of the sum rules see Supplemental Material \cite{SI}).

In Fig. \ref{fig5}b we show the Ti-XMCD spectra measured in the TEY-mode. Similarly to all the other LAO/ETO/STO q2DEGs investigated so-far \cite{Stornaiuolo2016,Chen2022}, the XMCD spectra cannot be reproduced by a positive exchange-interaction term. In particular, the XMCD spectral features are positive at L$_3$ and at L$_2$, a result which shows that Ti3d-states are characterized by a 3d-4f antiferromagnetic super-exchange interaction with Eu$^{2+}$ magnetic moments. The Ti-XMCD also show a finite integral, corresponding to a finite but small Ti-3d orbital moment of 0.005 $\mu_{B}$/Ti, with the same sign of the orbital moment measured in the Eu-XMCD data.

To investigate the influence of the FE polarization on the magnetic properties of the q2DEG, in Figs. \ref{fig5}c-d we show the Eu M$_{4,5}$-edge XAS and XMCD spectra acquired at 0.5 T, well below saturation, for the LAO/ETO/Ca(2\%)-STO heterostructure as a function of the FE polarization state, without (Fig. \ref{fig5}c) and with  (Fig. \ref{fig5}d) an applied back-gate voltage. In the remnant (V\textsubscript{gate} = 0) P$\downarrow$ and P$\uparrow$ states (Fig. \ref{fig5}c), a clear modulation of the XMCD spectra is observed, which becomes more pronounced under gate voltages of $\pm$80 V, sufficient to switch the polarization direction (Fig. \ref{fig5}d). The most prominent change is seen in the integral of the TFY-XMCD signal, which increases systematically from negative to positive gate voltage; equivalently, the XMCD intensity in the P$\uparrow$ state is higher than that in the  P$\downarrow$ state, where it is close to zero.

Fig. \ref{fig5}e presents the results of a full polarization-switching loop experiment, showing the evolution of both the spin moment and orbital/spin moment ratio vs. V\textsubscript{gate}. A clear hysteresis loop is observed, demonstrating full modulation of the Eu$^{2+}$ moments. While the change in the spin moment at 0.5 T is of the order of a few percent, the orbital moment varies by up to more than 50\%. Actually, the XMCD-integral (q, in Fig. \ref{fig5}c) is almost null in the P$\downarrow$, while it becomes clearly different from zero in the P$\uparrow$ configuration. These findings provide direct evidence of a non-volatile switching of the Eu$^{2+}$ orbital moment, i.e. an orbital magnetoelectric effect, as revealed by in-situ TFY-XMCD measurements under back-gating.

\section{Discussion and conclusions}
The experimental results presented in this work provide compelling evidence that the q2DEGs formed in LAO/ETO/Ca–STO heterostructures at the interfaces between FE–Ca–STO and the band-insulating oxides LAO and ETO exhibit switchable interfacial FE-dipoles. These systems therefore behave as FE-like two-dimensional metals, where the electronic ground state of the confined carriers is intimately coupled to the underlying FE order.

In addition, our results demonstrate a direct magnetoelectric coupling between the FE order parameter and the magnetic moments in the spin-polarized q2DEG formed at the LAO/ETO/Ca–STO heterostructure, establishing this system as a model multiferroic q2DEG, in which spin, orbital, and lattice degrees of freedom are interlaced and can be reversibly tuned by electric fields.

Two key experimental effects accompany the onset of ferroelectricity and its electrical switching:
(1) a non-volatile modulation of Ti$^{3+}$ occupation and a FE-driven modification of the Ti-3d orbital splitting;
(2) the emergence of a finite, electrically switchable, Eu-$M_{4,5}$ XMCD integral, i.e. a finite orbital magnetization.

The temperature and polarization dependence of the Ti-L$_{2,3}$ XLD spectra provide direct evidence that the FE-polarization strongly modulates the Ti$^{3+}$ anisotropic occupation and the Ti–O displacements at the interface. Below the FE-transition temperature, both LAO/Ca–STO and LAO/ETO/Ca–STO heterostructures display a clear enhancement of the XLD amplitude and a negative integral of the dichroic signal, indicating an increase of the 3d$_{xz,yz}$ orbitals occupation. This behavior is consistent with an increased Ti–O off-centering along the interface normal and reflects the emergence of a built-in interfacial FE dipole within the q2DEG region. Moreover, by quantitatively analyzing the XAS and XLD data by ligand-field atomic multiplet calculations we find that the FE transition and the FE-switching between P$\uparrow$ and P$\downarrow$ states induce variations of the orbital splitting and a departure from ideal octahedral coordination in the FE-state. This indicates that the FE-polarization modifies both the confining electric field and the local orbital occupancy within the metallic quantum well, giving rise to a non-volatile electric control of charge and orbital polarization in the q2DEG.

The FE-order of the q2DEG has a profound impact on the magnetic properties of the spin-polarized LAO/ETO/Ca-STO interface, giving rise to a clear non-volatile modulation of both the spin and orbital magnetic response of the heterostructure. This effect is directly revealed by the evolution of the Eu M$_{4,5}$ TFY-XMCD spectra across the FE hysteresis loop (Fig. \ref{fig5}).
Beyond the modulation of the Eu-spin moment, a pronounced change of the XMCD integral is observed, indicating the emergence of a non-trivial orbital magnetic contribution that is strongly coupled to the FE-polarization state. 

In the P${\downarrow}$ configuration, where transport and XLD data indicate that 3d$_{xy}$ sub-bands are occupied, the XMCD integral is nearly zero. This behavior is consistent with a quenched orbital moment, as expected for  Eu$^{2+}$ ions in the half-filled 4f$^{7}$ configuration. In contrast, in the P${\uparrow}$ state and upon increasing the positive back-gate voltage, thus the occupied 3d$_{xz,yz}$ states, the XMCD integral becomes distinctly negative, corresponding to an orbital magnetization parallel to both the spin moment and the applied magnetic field. Remarkably, this orbital contribution develops at low magnetic fields ( already at 0.1 Tesla), well below magnetic saturation, and reaches a sizable value at 0.5 T and at 6 T (i.e. in saturation) in the unpoled state where the FE-polarization is spontaneously P${\uparrow}$. Such a behavior is highly unusual for a nominally quenched orbital moment of Eu$^{2+}$ system and calls for a careful examination of the underlying microscopic mechanism.

Several possible mechanisms can be ruled out. First, structural distortion induced redistribution of Eu-4f states, i.e. an anisotropy of the 4f-orbitals associated to the interfacial symmetry breaking and to the FE-induced lattice deformation, is unlikely given the nature of Eu$^{2+}$ 4f shell and the large orbital to spin moment ratio (up to 0.1). As a matter of fact, XLD measurements at the Eu M$_{4,5}$-edge  (see Supplemental Material \cite{SI}) reveal a small, albeit finite, anisotropy, which is on the other hand anti-correlated to the XMCD integral: it is larger when the XMCD-derived orbital contribution is quenched (P${\downarrow}$) and decreases when the orbital signal is enhanced (P${\uparrow}$). This mechanism is clearly incompatible with the unquenched orbital moment.

Second, a pure van Vleck paramagnetic contribution from the minor Eu$^{3+}$ fraction can also be excluded. Such a mechanism would yield an orbital moment that is both too small in magnitude and opposite in sign to the experimentally observed one. On the other hand, invoking a strong (negative) exchange interaction with Eu$^{2+}$ spin to enhance the Eu$^{3+}$ orbital moment and to get the right sign, would require a sizable and clearly identifiable Eu$^{3+}$ XMCD signal, with the appropriate sign and magnitude, which is not observed (see Supplemental Material \cite{SI}). Furthermore, this scenario also fails to explain the pronounced gate-voltage dependence, which primarily modifies the splitting and anisotropic occupation of the Ti-3d q2DEG states.

\begin{figure}[ht!]
	\includegraphics[width=90mm]{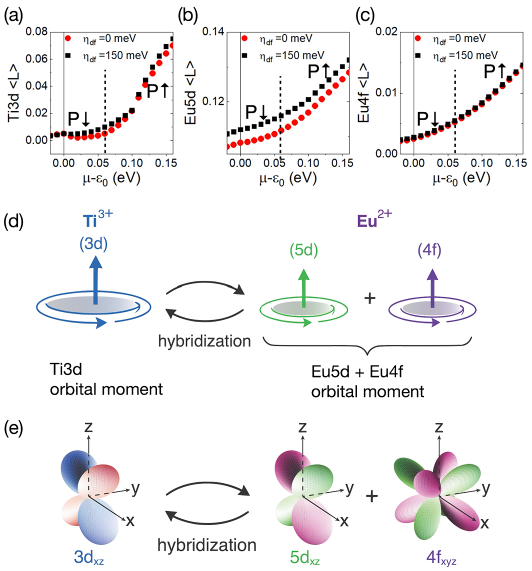}
	\caption{\label{fig6} Calculated orbital angular momentum  (in $\mu_B$/Eu units) as function of the $\mu$-$\epsilon_{0}$ ($\mu$ the chemical potential, $\epsilon_{0}$ the bottom of the conduction band) of (a) Ti3d, (b) Eu5d and (c) Eu4f states. Red circles are calculated without Eu5d-Eu4f hybridization, while black square data are calculated for $\eta_{df}$=150 meV. The chemical potential roughly corresponding to P${\downarrow}$ and P${\uparrow}$ in the figures is indicated in each panel. The dashed line marks the Lifshitz point according to theory. (d) and (e) A sketch of the mechanism leading to the transfer of Ti3d orbital momentum to Eu5d and Eu4f states.}
\end{figure}

Having excluded experimental artifacts as well as structural and trivial ionic explanations, the origin of the finite Eu XMCD magnetic orbital signal and its dependence on the FE-polarization require consideration of the electronic structure of the LAO/ETO/STO heterostructure. 


Ab-initio calculations on (001) LAO/ETO/STO interfaces \cite{DiCapua2022, Wadhwa2023} showed that Eu-5$d$ states possess a finite spectral weight at the Fermi level and are strongly hybridized with both Eu-4$f$ and Ti-3$d$ orbitals. In this framework, the Ti-3$d$ carriers forming the q2DEG become spin-polarized through exchange coupling to Eu$^{2+}$ 4f magnetic moments, mediated by Eu-5$d$ states. Moreover, valence band spectroscopy on (001) LAO/ETO/STO shows that the Eu-$4f^{7}$ VB peak, located around 2 eV below the Fermi level, has a very long tail almost merging with the conduction band features \cite{DiCapua2022}. This suggests that Eu-$4f^{7}$ states hybridize to some extent with states at the Fermi level, including Ti-3$d$ and Eu-5$d$ states.

Thus, to identify the microscopic origin of the electrically switchable Eu orbital moment, we developed a multi-orbital tight-binding model based on Wannier functions derived from density-functional theory, which includes the Ti-3$d$, Eu-5$d$, Eu-4$f$, and O-2$p$ orbitals, together with their mutual hybridization, local Coulomb interactions, and atomic spin-orbit coupling (see Supplemental Material for details \cite{SI}). This framework simultaneously captures the spin-polarized q2DEG and the hybridization between the Ti and Eu electronic states that is characteristic of the LAO/ETO/STO interface. 
Two microscopic ingredients emerge as essential for understanding the experimental results. First, electrostatic gating modifies the occupation of the Ti-3$d$ sub-bands and drives the system through the Lifshitz transition, producing a strong increase of the orbital magnetization associated with the itinerant q2DEG. Second, inversion-symmetry breaking at the interface allows hybridization between Eu-5$d$ and Eu-4$f$ orbitals, providing a channel through which this orbital magnetization can be transferred to the Eu ions. The calculated orbital moments are summarized in Fig. \ref{fig6}. 
In the low-density regime, where only the Ti-$d_{xy}$ band is occupied, the orbital moments projected onto both Ti-3$d$ and Eu-5$d$ orbitals are small. Once the Ti-$d_{xz,yz}$ bands become populated, both moments increase rapidly and exhibit nearly identical carrier-density dependence, indicating their common origin in the orbital polarization of the itinerant electrons. The larger Eu-5$d$ moment mainly reflects the stronger spin-orbit coupling of the 5$d$ shell.

In contrast, even after including inversion-breaking hybridization described by the effective parameter $\eta_{df}$, the calculated Eu-4$f$ orbital moment remains very small ($<0.02\mu_B$/Eu), far below the experimentally observed value, demonstrating that hybridization alone cannot account for the XMCD results.

To explain the experimental magnitude, we introduce a minimal configuration-interaction model involving the two lowest-energy Eu configurations, $4f^7$ and $4f^65d^1$. The Hamiltonian includes the local 4$f$-5$d$ hybridization and an effective Coulomb interaction between the two shells,
\begin{equation}
H = H_{4f} + H_{5d} + H_{\rm hyb} + H_{fd},
\end{equation}
where $H_{\rm hyb}$ describes the inversion-induced mixing between 4$f$ and 5$d$ orbitals, and $H_{fd}$ accounts for the effective orbital interaction due to the Coulomb coupling (see Supplemental Material \cite{SI} and ref. \cite{Georges2013}). 

Because the $4f^7$ ground state has essentially zero orbital angular momentum, the Eu orbital moment originates from the admixture with the orbital active $4f^65d^1$ configuration. The Coulomb interaction lowers the energy of this configuration when the 4$f$ and 5$d$ orbital moments are aligned, thereby increasing its weight in the ground state. 
As shown in the Supplemental Material \cite{SI}, this gives rise to an orbital moment of Eu-$4f$ states  \begin{equation}
\mu_f = \frac{3\mu_B}{2}
\left[ 1- \frac{\Delta_{eff}} {\sqrt{(\Delta_{eff})^2+4\eta_{df}^2}}
\right],
\label{eq:orbitalmoment}
\end{equation}
where $\eta_{df}$ describes the local Eu-$5d$/Eu-$4f$ mixing, which incorporates both the inversion-symmetry breaking crystal field and the structural distortions of the Eu-O and Ti-O octahedra, and $\Delta_{eff}$ is the effective energy separation between the $4f^6 5d^1$ and~$4f^7$ configurations, rescaled by the orbital interaction due to the Coulomb term.

According to the experimental results, the FE-polarization acts on the Eu-$5d$/Eu-$4f$ local hybridization and on the effective energy separation. In P${\downarrow}$ configuration, and in particular for negative electrostatic gating, we can assume $\eta_{df}\simeq 0$, which corresponds to $\mu_{f}\simeq 0$. Conversely, in P${\uparrow}$ configuration and with positive gating, a finite hybridization of $\eta_{df}\simeq 0.2 eV$ together with reasonable values of $\Delta_{eff}$ between 0.4 and 0.6 eV will give rise to  an orbital moment for Eu-$4f$ in the 0.25 to 0.4 $\mu_B$/Eu range, in reasonable agreement with the experiment. 

While Eu-$M_{4,5}$ XMCD probes a local quantity and cannot directly measure the total orbital magnetization of the q2DEG, the combined experimental and theoretical results consistently support a picture in which ferroelectric polarization controls the orbital character of the itinerant electron gas, and its equilibrium orbital moment. This orbital magnetization is transferred to Eu-5$d$ states through Ti-3$d$/Eu-5$d$ hybridization and subsequently projected onto the Eu-4$f$ shell through local 4$f$-5$d$ coupling, making the Eu orbital response a spectroscopic fingerprint of the orbital magnetization developing within the q2DEG. Similar enhanced and unusual orbital magnetization, with an m$_{orb}$/m$_{spin}$ of the order of 0.1, has been
detected in 2D Rashba ferromagnetic systems, like VI$_{3}$ \cite{hovancik2023}. 

These findings are consistent with theoretical expectations that orbital moment in inversion-asymmetric oxide q2DEGs is strongly enhanced by multiband occupancy. In particular, recent theoretical and experimental studies have highlighted the presence of large orbital textures and orbital angular momentum in Rashba-split oxide electron systems, as evidenced by the prediction and observation of the orbital-Rashba Edelstein effect \cite{Johansson2021,ElHamdi2023}. Within this framework, the occupation of Ti-$d_{xz,yz}$ states is expected to substantially enhance the orbital moment of the electronic states near the Fermi level. Our calculations indicate that this enhanced orbital moment is efficiently transferred to Eu-5$d$ states through hybridization and can subsequently induce a finite orbital response of the Eu-4$f$ shell.

Altogether, these findings demonstrate that ferroelectricity and magnetism are deeply intertwined in oxide q2DEGs formed at Ca-STO based interfaces. The ferroelectric polarization controls both the charge localization and orbital structure of Ti-3$d$ states, while simultaneously modulating the Eu orbital response through interfacial hybridization. The coexistence and coupling of ferroelectric, ferromagnetic, and orbital degrees of freedom establish LAO/ETO/Ca-STO as a prototypical multiferroic two-dimensional electron gas, providing a versatile platform for exploring electrically tunable spin--orbit-orbital interactions and for developing future oxide-based quantum devices.

\begin{acknowledgments}
M.S., D.S., M. B. and Y. C. acknowledge support from the European Union's Horizon Europe project IQARO (GA n. 101115190), and from the MUR (Italian Ministry of University and Research) within the projects PRIN PNRR 2022 SONATA (No. P2022SB73K), PRIN 2022 STIMO (No. 2022TWZ9NR) and PNRR Project PE0000023-NQSTI, funded by the European Union – NextGenerationEU. DP acknowledges the ANR-France (French National Research Agency) for its financial support of the T-ERC project ORBIFUN ANR-23-ERCS-0003-001. M. R. acknowledges support from the European Union's Horizon Europe research and innovation programme within the Marie Sklodowska-Curie project MODERN (GA No. 101108695). X-ray absorption spectroscopy measurements were collected at the XTreme beamline of the Swiss Light Source facility during experiment/proposal number 20222194.

\end{acknowledgments}

\appendix




\nocite{*}


%

\end{document}